\documentclass[aps,prl,reprint,superscriptaddress,showpacs,nobibnotes]{revtex4-1}
\usepackage{graphicx}
\usepackage{amssymb}
\usepackage[utf8]{inputenc}
\usepackage{amsmath}
\usepackage{color}
\usepackage[normalem]{ulem}
\usepackage[english]{babel}

\renewcommand{\a}{\alpha}

\newcommand{\KK}{\mathbf{K}}
\newcommand{\CC}{\mathbf{C}}
\newcommand{\EE}{\mathbf{E}}
\newcommand{\HH}{\mathbf{H}}

\renewcommand{\S}{\mathbf{\Sigma^{(\mathbf{x})}}}

\newcommand{\SSS}{\mathbf{\Sigma}}
\renewcommand{\r}[1]{\:\mathbf{r(}#1\mathbf{)}}
\newcommand{\xx}{\mathbf{x}}

\newcommand{\x}[1]{\mathbf{x}({#1})}

\newcommand{\m}{\boldsymbol\mu}

\newcommand{\Id}{\mathbb{I}}

\begin{document}

\title{Diversification versus specialization -- lessons from \\ a noise driven 
linear dynamical system}

\author{Gabriell M\'at\'e}
\affiliation{Heidelberg University, Institute for Theoretical Physics,  Heidelberg, Germany}
\author{ Z. N\'eda}
\affiliation{Babe\c{s} - Bolyai University, Department of Physics,  Cluj-Napoca, Romania}
\affiliation{Edutus College,Tatab\'anya, Hungary}


\begin{abstract}
Specialization and diversification are two major strategies that complex systems might exploit. 
Given a fixed amount of resources, the question is whether to invest this in elements 
that respond in a correlated manner to external perturbations, or to build a diversified system with 
groups of elements that respond in a not necessarily correlated manner. 
This general dilemma is investigated 
here using a high dimensional discrete dynamical system subject to an external noise, analyzing the statistical properties of an order parameter that quantifies growth.  
Our analytical solution suggests that diversification is a good strategy once the
system has a fair amount of resources. For systems with small or extremely large supplies, 
we argue that specialization might be a more successful strategy. We discuss the results also 
from the perspective of economic and biologic systems.
\end{abstract}
\maketitle

Many physical systems are in a first approximation linear \cite{tomas-rodriguez_linear_2010}. Well-know examples are electric circuits with simple elements, mechanical strain-stress relations, transport phenomena or kinetics in chemical reactions. These systems often lead to iteratively repeating dynamics. Even in high dimensional cases,  the dynamics of linear systems are easy to predict \cite{curtain_introduction_1995}. In case however they are subjected to an external noise, their dynamical response and consecutively their statistical properties might become interesting \cite{soderstrom_discrete-time_2002}. The observed processes may be similar to the ones perceived in a variety of other, not necessarily physical systems \cite{steinbrecher_generalized_2004,eldar_functional_2010,silver_neuronal_2010,assaf_extrinsic_2013}. Here we plan to address the problem in a more general, interdisciplinary context, using a simple, analytically solvable model.  Particularly, we are interested in comparing the statistical properties of linear systems composed entirely by positively 
correlating elements with that of systems that have elements that anti-correlate with each other.  We argue that the former one
can be viewed as a complex system using a specialization strategy, while the latter one would correspond to a
system that exploits the advantage of diversification. 

We can put now the problem in a more clear interdisciplinary context, quoting, for instance, the golden rules of portfolio optimization, "don't pull all your eggs in the same basket''.  Investors have been applying the principle of diversification for millennia, the concept being mentioned also in the Bible (Ecclesiastes 11:2). With the emerging of modern portfolio theory different algorithms were developed to assemble a group of diverse assets which satisfies certain predefined criteria with respect to the perceived risks and the expected returns \cite{plerou_random_2002,hagin_investment_2004,bera_optimal_2008,wang_multifactor_2009}. The topic however is highly debated when it comes to managing economic entities. It is still not clear, for instance, whether the future of a company is safer if the company attempts to diversify its activities or if it aims to specialize them \cite{kenny_diversification_2009}.  In the same manner, biological evolution or social behavior is also influenced by this twofold way. 
 
Naturally, there are many influencing factors and many paths to success, however the natural tendency of systems to diversify on their own seem to have at least an evolutionary advantage. Genetic diversity and the genetic variability of a species enables its adaptation to fast environmental changes, avoiding extinction \cite{lande_role_1996,elena_evolution_2003,de_aguiar_moran_2011,rulands_specialization_2014,korolev_genetic_2010,meyer_effects_2007}. 

The question regarding diversification also extends to choosing the economic strategies of countries. Arguments and empirical data are conflicting when it comes to answering the question whether a country should specialize or should diversify it's industry. The insightful study of Imbs and Wacziarg \cite{imbs_stages_2003} indicates that stages of diversification heavily depend on the level of development of the respective countries: while emerging and developing countries are usually highly specialized, the developed ones tend to diversify their economy. On the other end of the spectrum, the richest countries have a tendency to specialize again. 

The message of diversification is clear, the principle is applied to reduce risks \cite{peters_ergodicity_2013}. Or, turning the problem around, diversification is the result of a self-organization which tunes a system to hedge risks. 
While the principle seems inherent, in most cases it is hard to grasp it in a rigorous scientific way. Also, there must be a limit to the benefits of diversification with respect to the size of the system. In fact, this can be seen in biological systems, wherein, given a very small population, the random process of genetic mutation and drift overwhelm natural selection \cite{elena_evolution_2003}.

If diversification is omnipresent in nature, how is it possible that certain systems, especially economic systems seem to perform better if they specialize? Is the principle of competitive advantage \cite{porter_competitive_2011}, which favors the specialization of the industries of a nation, in contradiction with diversification? Why do developed nations diversify their economy? Is there a universal explanation for the system-size limit below which the principle of diversification is not valid? 
 These are indeed questions that worth to be answered through a general approach, validating the 
 the ideas formulated by Imbs and Wacziarg \cite{imbs_stages_2003}. Here, we offer insights on strictly mathematical basis, concluding some results obtained form stochastically driven linear dynamical systems.  
  
Let us consider a dynamical system composed of $M$ elements. At time $t$, the state of each element $i$, $i \in \{1..M\}$, can be characterized with a continuous variable, $r_i(t)$. This variable can and will have a different meaning in each real-world scenario. For instance, it can represent a measurable physical quantity, the value of assets for a company, economic indexes for different countries, or it can be a hypothetical quantity characterizing the comfort of an individual in a population.  We assume a very simple discrete time-evolution law. Each element $i$ is influenced by an external environmental effect and the previous states of the other elements in the system. Mathematically this is expressed by the following equation:
\begin{equation}
	\r{t+1} = \CC \r{t} + \mathbf{f(x'}(t)\mathbf{)} \mbox{,}
	\label{eq_prototype0}
\end{equation}
where $\r{t}$ is the $M$ dimensional vector composed of all the $r_i$ state variables. The first term on the right side stands 
for the interaction between the actors, the $M \times M$ matrix, $\CC$,  defines the strength of the interaction/coupling between the actors. $\mathbf{x'}$ is an $M$ dimensional vector, representing the state of the local environments of the elements ( this can be for instance, local nutrition concentration, the mood of the market for a given product, etc.). $\mathbf{f}$ is a vector valued function which maps the $\mathbf{x'}$ state of the environments to the influence these have on the state variables 
$\mathbf{r}$. 

To study such a model, we need to explicitly specify $\mathbf{f}$. As this is a very difficult task on it is own, we will work in a first order approximation, and assume that the environment has a linear effect on the state variables. While this obviously cannot be true on very long time scales, we can make this approximation assuming short enough time-steps, and neglect thus higher order terms. Equation (\ref{eq_prototype0})  writes then in the following form:
\begin{equation}
	\r{t+1} = \CC \r{t} + \xi \mathbf{x'}(t) \mbox{,}
	\label{eq_prototype}
\end{equation}
where $\xi$ is a constant. The equations written here are very similar with the famous 
evolution equations in the quasi-species theory for a system of information carrying macromolecules,
as it was formulated by M. Eigen \cite{eigen}. Also, such dynamics are appropriate to describe the one-dimensional diffusion of particles that are subject to a generalized "flocking"-type coupling mechanism. Depending on the entries of the $\CC$ matrix, particles may tend to follow each others' movement or to consider an opposite changes in their coordinates. 

To be able to handle this equation analytically, further simplifications are needed. First we assume that the different elements 
composing our system are subjected to randomly changing environments which are similar in nature, and that the effect of these environmental factors are distributed according to a normal distribution. In such case, $\xi$ can be absorbed into these random variables, and eq. (\ref{eq_prototype}) writes as 
\begin{equation}
	\r{t+1} = \CC \r{t} + \x{t} \mbox{,}
	\label{eq_prototype}
\end{equation}
where $\x{t}$ is distributed according to a multivariate normal distribution with $M$ dimensional mean vector $\m$ and $M \times M$ covariance matrix $\S$.  We denote this as: 
\begin{equation}
	\x{t} \sim N(\m, \S)  \mbox{.}
\end{equation}
Naturally, $\x{t}$ and $\x{t+1}$ are independent and identically distributed. Furthermore, we require 
$\CC$ to be symmetric and $\CC^2$ to be positive semidefinit. 

We define the initial conditions as $\r{0}=0$, and solve eq. (\ref{eq_prototype}) recursively using the properties of affine transformations performed on multivariate normal distributions \cite{tong_multivariate_2011}. The generic solution gives:
\begin{align}
  \r{t} &\sim N\left(\sum_{k=0}^{t-1}\CC^k \m, \sum_{k=0}^{t-1}\CC^k \S \CC^k \right) \label{eqaltalanos0}\mbox{.}
\end{align}

We are interested now on the differences in the statistical properties of $\r{t}$ that might appear due to diversification. Therefore, we will study two distinct cases: the completely \textit{homogeneous system} in which all elements of the system behave in a correlated manner,  and the \textit{inhomogeneous case} where diversity appears due to the presence of anti-correlated elements in the system. In the homogeneous case the system is not diverse, there are only positive couplings between the elements, meaning they all benefit from each-other's ``well being'', also, they all loose by the others misfortune. This means that the entries of the $\CC$ matrix are all positive. On the other hand in our inhomogeneous system there are two ``types'' of elements.  We do not specify what the word "type'' means here, we only indicate that elements of the same type benefit from each other's well-being and they are drawn back by the well-being of elements of the other type. This means that the $M$ elements of the system are separated in two blocks so that the $\CC$ matrix coefficients corresponding to elements in the same block are positive, and the $\CC$ matrix coefficients corresponding to elements in different blocks are negative. Note that since $\CC^2$ will have the same block structure as $\CC$, for cases of $\S$ which do not change this structure, the negative blocks in $\CC$ will result in anti-correlation between the corresponding elements. In the same fashion, a homogeneous $\CC$ will yield only positive correlations in the system.

For the homogeneous system we consider the simplest choice for $\CC$, where all the entries except the diagonal are values $\alpha>0$.  The diagonal will contain values $l$ such that $l>\alpha$, as we assume that states of the individual elements composing the system should depend more on their previous state rather than the previous state of the other parts of the system. For the inhomogeneous system $\CC$ will have a block form, with two homogeneous blocks of $\alpha$, while the entries corresponding to the coupling between different type of elements will be $-\alpha$. Similarly to the homogeneous case, diagonal elements will be set to a positive value $l>\alpha$. 

In the homogeneous case, we can write thus the $\CC$ matrix as:
\begin{equation}
 \CC = \alpha \EE_M + (l-\alpha)\Id_M
\end{equation} 
where $\EE_M$ is  the $M \times M$ matrix with all entries one
and $\Id_M$ is the $M \times M$ identity matrix. Thus, $\CC$ has entries of $l$ along its diagonal.

In the inhomogeneous case, we can give the matrix $\CC$ in the form:
\begin{equation}
\CC =
 \begin{pmatrix}
  \alpha \EE_{M/2} & -  \alpha \EE_{M/2}  \\
  - \alpha \EE_{M/2} &  \alpha \EE_{M/2}
  \end{pmatrix} + (l-\alpha) \Id_M = \a \HH_M  + (l-\a) \Id_M 
\end{equation}
While formally this notation assumes that $M$ is even, we don't necessarily need to divide the system into two equally sized subsystems. However, for simplicity we will carry on with this form. 
$\HH_M$ denotes the $M \times M$ matrix with two blocks of $1$ along the diagonal and with elements of $-1$ outside these blocks.

Then, we can rewrite eq. (\ref{eq_prototype}) for both of these cases as 
\begin{equation}
	\r{t+1} = \a \KK \r{t} + \x{t} \mbox{,}
\end{equation}
with $\KK=\EE_M + (\lambda-1) \Id_M$ in the homogeneous case, $\KK = \HH_M+ (\lambda-1) \Id_M$ in the inhomogeneous case, where $\lambda=l/\alpha>1$. 

The generic solution of Eq. (\ref{eqaltalanos0}) writes than as 
\begin{align}
  \r{t} &\sim N\left(\sum_{k=0}^{t-1}(\a \KK)^k \m, \sum_{k=0}^{t-1}\a^{2k} \KK^k \S \KK^k \right) \label{eqaltalanos}\mbox{.}
\end{align}

We assume now that the elements of the $\xx$ external noise are uncorrelated and identically distributed. In such case
\begin{eqnarray}
\m = \mu
 \begin{pmatrix}
  1 & 1 & 1 &\cdots 1 \\
  \end{pmatrix}^T \mbox{,} \\
\S = \sigma^2 \mathbb{I} \mbox{,}
\end{eqnarray}
where $\mu$ denotes the expected value and $\sigma$ the standard deviation of the normally distributed noise components.
The superscript $T$ denotes the transpose of a matrix.

Solution (\ref{eqaltalanos}) rewrites in such case as:
\begin{align}
  \r{t} &\sim N\left(\sum_{k=0}^{t-1}(\a \KK)^k \m, \sum_{k=0}^{t-1}(\a \KK)^{2k} \sigma^2 \right) \label{eqspec0}\mbox{.}
\end{align}

To proceed further, we need to find the expressions for $(\a \KK)^k$ and $(\a \KK)^{2k}$.  In order to do this we consider 
separately the homogeneous and inhomogeneous case.

In the homogeneous case taking into account that 
$\Id^n = \Id$ and $\EE_M^n=M^{n-1} \EE_M$  a simple algebra leads us to
\begin{equation}
\KK^k=\delta^k \mathbb{I}_M+\frac{\EE_M}{M} \left( Q^k-\delta^k \right),
\end{equation}
where $\delta=\lambda-1>0$, and $Q=M+\lambda-1$.

For the $t \rightarrow \infty$ limit and $\alpha Q <1$  it is straightforward to show now that:
\begin{eqnarray}
\label{sum1h}  \sum_{k=0}^\infty (\alpha \KK)^k=\frac{\mathbb{I}_M}{1-\alpha \delta} +  \frac{\alpha \EE_M }{(1-\alpha Q)(1-\alpha \delta)}  \\
\label{sum2h} \sum_{k=0}^\infty (\alpha \KK)^{2k}=\frac{\mathbb{I}_M}{1-\alpha^2 \delta^2} + \frac{\alpha^2 \EE_M (Q+ \delta)}{(1-\alpha^2Q^2)(1-\alpha^2 \delta^2)}
\end{eqnarray}

In case of $\alpha Q \ge 1$ the series are diverging, and thus the $\r{t}$ values are not converging to any finite value. In the following we will be interested only the cases where $\alpha Q<1$, and the system has statistically stable
properties.

In the inhomogeneous case  we observe now, that $\HH_M^n=M^{n-1} \HH_M$, and we get an expression similar to the homogeneous case:
\begin{equation}
\KK^k=\delta^k \mathbb{I}_M+\frac{\HH_M}{M} \left(Q^k-\delta^k \right)
\end{equation}

Taking the $t \rightarrow \infty$ limit in the $\alpha Q<1$  case one gets the same results as in 
eq. (\ref{sum1h}) and (\ref{sum2h}), with $\EE_M$ replaced by $\HH_M$. 

We can view now the advantages and/or disadvantages of the diversity that our model inhomogeneous system presents from different perspectives. The first possibility is to consider the problem from the view of one element (or actor) in the 
composed system and another possibility is to consider it from the perspective of the whole ensemble.  

We will focus on the average value and standard deviations of $\r{t}$ observed in these two perspectives. 
These quantities can be calculated using simple affine transformations of the form $y = \mathbf s\r{t}$. 
For such affine transformations it is known that:
\begin{equation}
	y \sim N \left( \mathbf s \m_{\r{t}},  \mathbf s \SSS^{(\r{t})} \mathbf s^T \right) \mbox{.}
\end{equation}

If we are interested to compute the averages ($\mu^{(1)}$)  and standard deviation  ($\sigma^{(1)}$) experienced by one element in the system, we choose for  $\mathbf s$ the $M$ dimensional vector
\begin{equation}
\mathbf{s}^{(1)} =
 \begin{pmatrix}
  1 & 0 & 0 &\cdots 0 \\
  \end{pmatrix} \mbox{.}
\end{equation}
In such case, it can be shown that:
\begin{eqnarray}
\mathbf{s}^{(1)} \mathbb{I}_M \m=\mu; \: \:
\mathbf{s}^{(1)} \EE_M \m=M {\mu}; \\
\mathbf{s}^{(1)} \HH_M \m=0; \: \:
\mathbf{s}^{(1)} \mathbb{I}_M \mathbf{s}^{(1)T} =1; \\
\mathbf{s}^{(1)} \EE_M \mathbf{s}^{(1)T}=1; \: \:
\mathbf{s}^{(1)} \HH_M \mathbf{s}^{(1)T}=1 
\end{eqnarray}

In the homogeneous case (labeled from now on with $h$) using equation (\ref{eqspec0}) with equations (\ref{sum1h}) and (\ref{sum2h}) we get:
\begin{eqnarray}
\label{m1h}
\mu_h^{(1)}=\frac{\mu}{1-\alpha Q} \\
\label{s1h}
\sigma_h^{(1)}=\sigma \sqrt{\left [ \frac{1}{1-\alpha^2 \delta^2}+ \frac{\alpha^2(Q+\delta)}{(1-\alpha^2Q^2)(1-\alpha^2 \delta^2)} \right ]}
\end{eqnarray}

In the inhomogeneous case (labeled from now on with $ih$), we would have:
\begin{eqnarray} \label{m1i}
\mu_{ih}^{(1)}=\frac{\mu}{1-\alpha \delta} \\
\label{s1i} \sigma_{ih}^{(1)}=\sigma_h^{(1)}
\end{eqnarray}

In the inhomogeneous case, we obtain thus that the expected value for one element in the system is independent
of the system size. In contrast with this, in the homogeneous case the expected value for one element 
in the system increases in modulus as the system size increase. This is good when $\mu >0$, but inconvenient
when $\mu<0$. Elements in large homogeneous systems are rather 
vulnerable for negative external effects. This vulnerability is much reduced in an inhomogeneous structure.  
When $\mu \ne 0$ 
from equations (\ref{m1h}) and (\ref{m1i}), we get that
\begin{equation}
\frac{\mu_h^{(1)}}{\mu_{ih}^{(1)}}= \frac{1-\alpha \delta}{1-\alpha Q} >1 \mbox{,}
\end{equation}
which means that the expected value will always be smaller in the inhomogeneous case. 
In case of $\mu=0$, one gets $\mu_h^{(1)}=\mu_{ih}^{(1)}=0$. In the limit $\alpha \ll 1$, we get that the ratio is of first order in $\alpha$:
\begin{equation}
\frac{\mu_h^{(1)}}{\mu_{ih}^{(1)}}\approx 1+\alpha M > 1 
\end{equation}

From equations (\ref{s1h}) and (\ref{s1i}), it results that for the same system size the standard deviation is the same.  One can also easily prove that for $\alpha <1/Q$ the standard deviation experienced by one element is increasing with increasing system size. 

Let us view now the homogeneous and inhomogeneous system in the viewpoint  of the whole ensemble, defining a value of $r$ averaged on all actors. This can be done by: 
 \begin{equation}
 \mathbf{s}^{(M)} =\frac{1}{M}
 \begin{pmatrix}
  1 & 1 & 1 &\cdots 1 \\
  \end{pmatrix} \mbox{.}
\end{equation}

Straightforward calculations lead to:
\begin{eqnarray}
\mathbf{s}^{(M)} \mathbb{I}_M \m=\mu; \: \:
\mathbf{s}^{(M)} \EE_M \m=M {\mu}; \\
\mathbf{s}^{(M)} \HH_M \m=0; \: \:
\mathbf{s}^{(M)} \mathbb{I}_M \mathbf{s}^{(M)T} =1/M; \\
\mathbf{s}^{(M)} \EE_M \mathbf{s}^{(M)T}=1; \: \:
\mathbf{s}^{(M)} \HH_M \mathbf{s}^{(M)T}=0
\end{eqnarray}

In the homogeneous case
\begin{eqnarray} \label{mMh}
\mu_h^{(M)}=\mu_h^{(1)}=\frac{\mu}{1-\alpha Q} \\
\label{sMh} \sigma_h^{(M)}=\frac{\sigma}{ \sqrt{ M \left( 1-\alpha^2 Q^2 \right)}} \mbox{,} 
\end{eqnarray}
and in the inhomogeneous case:
\begin{eqnarray}
\label{mMi} \mu_{ih}^{(M)}=\frac{\mu}{1-\alpha \delta} \\
\label{sMi} \sigma_{ih}^{(M)}=\frac{\sigma}{ \sqrt{ M ( 1-\alpha^2 \delta^2 )}}.
\end{eqnarray}

Comparing the homogeneous and inhomogeneous systems, we get: 
\begin{equation}
\frac{\mu_h^{(M)}}{\mu_{ih}^{(M)}}=\frac{\mu_h^{(1)}}{\mu_{ih}^{(1)}}= \frac{1-\alpha \delta}{1-\alpha Q} >1
\end{equation}
For $\mu=0$, we get $\mu_h^{(M)}=\mu_{ih}^{(M)}=0$. In the limit of $\alpha \ll 1$ we get again an effect which is first order in
$\alpha$:
\begin{equation} \label{gain-whole-mu}
\frac{\mu_h^{(M)}}{\mu_{ih}^{(M)}}=\frac{\mu_h^{(1)}}{\mu_{ih}^{(1)}} \approx 1+\alpha M >1
\end{equation}

This means, that also from the viewpoint of the whole system, the expected value is always larger in modulus for the homogeneous case. In other words, inhomogeneous systems are less vulnerable as a whole. The vulnerability of homogeneous systems increase with their sizes. 

For the standard deviations we get:
\begin{equation}
\frac{\sigma_h^{(M)}}{\sigma_{ih}^{(M)}}= \sqrt{\frac{ 1-\alpha^2 \delta^2}{ 1-\alpha^2Q^2 }} >1
\end{equation}

In the limit of $\alpha \ll 1$, this yields
 \begin{equation}\label{gain-whole-sigma}
	\frac{\sigma_h^{(M)}}{\sigma_{ih}^{(M)}} \approx 1+\frac{\a^2 M^2}{2} > 1 \mbox{.}
\end{equation}
 
The standard deviation for the averaged $\r{t}$ value is thus also smaller for the inhomogeneous system, although 
the effect is only second order in the $\alpha M<1$ parameter. Inhomogeneity offers thus a kind of inertia against 
changes in the fluctuations of the global welfare of the system. This effect is again more prominent for larger systems. 
 
In order to discuss the obtained results in the perspective of an evolving a complex system, we have to take into account additional factors. First, any complex system is small at its ``birth''. This means that it operates with very limited resources. In a fair game, where the external noise has a zero expectation value, the game should remain statistically even over time, independently of the chosen strategy (e.g. the choice with respect to diversification and specialization). However, due to the fact that the initial resources are finite, the game is by no means "fair''. System resources cannot become negative (or, in some cases, deeply negative), since in such case the player is out of the game. Small systems do not have enough resources that could be divided in a diversified structure of elements with reasonable large startup values which would grant a good chance of surviving in a changing environment. In such scenarios, the good strategy is to specialize, and gamble the available resources in one direction. Then, odds are good that the resources will hold out and in case of favorable external factors they will grow rapidly. 
We should also recall now eq. (\ref{gain-whole-mu}) and (\ref{gain-whole-sigma}), which indicate that for a system composed by a small number of elements ($M$) the gain in diversification is not obvious. 

Once (or if) the total resources are increased, the system might like to insure its well-being by diminishing further risk. In such sense, the good strategy is to diversify the resources by creating a less vulnerable structure with non-correlated or anti-correlated elements. Then, the standard deviation of the "well being" parameter is decreased. At the same time, the statistical expectation value is also reduced as a trade-off. While this means that in favorable conditions the growth rate is reduced, under unfavorable conditions losses are limited by the same effect. As the portfolio is bigger and more diversified, the system as a whole presents a larger inertia against external perturbations, allowing a more stable growth if conditions are favorable. 

Further on, if the system in its evolution will have large resources, specialization becomes again a reasonable strategy. In this case the elements of the system will posses enough resources which enable survival while losing some assets in unfavorable conditions. On the other hand the system exploits the higher average gains during the favorable times. 

The discussed scenarios are in good agreement and explain the findings presented in reference \cite{imbs_stages_2003}. 
Being speculative, we might state that this process of shifting from specialization to diversification and back can be viewed as an eternal cycle, repeating itself many times in human and natural history.
   
From our model, we can also see, that small systems do not benefit from diversification. In case $\a M \ll 1$, our results
suggest that the standard deviation and the average values are rather similar in the homogeneous and inhomogeneous 
construction. It has to be mentioned however, that here $\a$ is a constant characterizing each particular system, and it's not universal.

Our final conclusion is that the decreased average value in an inhomogeneous construction is an important gain, and this represents a shield against negative influences. The damping effect introduced by the inhomogeneous structure will diminish the chance of getting in bankruptcy. Inhomogeneous communities, inhomogeneous resources in biological systems or an inhomogeneous portfolio used by brokers have in this perspective the same risk reducing advantages. On the other end of the spectrum, specialized 
systems offer quicker gains and may outperform diverse systems, granting higher returns. Of course, if a variety of resources is available, systems can specialize in certain aspects and diversify in others. However, the consensus is the same, the diversified part of these systems will be less vulnerable for unexpected events and the specialized part will be more profitable 
under favourable trends.

This research was supported by the European Union and the State of Hungary,
co-financed by the European Social Fund in the framework of T\'AMOP 4.2.4.A/2-11-1-2012-0001 
‘National Excellence Program’. Discussions with Prof. C. Gross are gratefully acknowledged. 


\begin{thebibliography}{1}
\makeatletter
\providecommand \@ifxundefined [1]{%
 \@ifx{#1\undefined}
}%
\providecommand \@ifnum [1]{%
 \ifnum #1\expandafter \@firstoftwo
 \else \expandafter \@secondoftwo
 \fi
}%
\providecommand \@ifx [1]{%
 \ifx #1\expandafter \@firstoftwo
 \else \expandafter \@secondoftwo
 \fi
}%
\providecommand \natexlab [1]{#1}%
\providecommand \enquote  [1]{``#1''}%
\providecommand \bibnamefont  [1]{#1}%
\providecommand \bibfnamefont [1]{#1}%
\providecommand \citenamefont [1]{#1}%
\providecommand \href@noop [0]{\@secondoftwo}%
\providecommand \href [0]{\begingroup \@sanitize@url \@href}%
\providecommand \@href[1]{\@@startlink{#1}\@@href}%
\providecommand \@@href[1]{\endgroup#1\@@endlink}%
\providecommand \@sanitize@url [0]{\catcode `\\12\catcode `\$12\catcode
  `\&12\catcode `\#12\catcode `\^12\catcode `\_12\catcode `\%12\relax}%
\providecommand \@@startlink[1]{}%
\providecommand \@@endlink[0]{}%
\providecommand \url  [0]{\begingroup\@sanitize@url \@url }%
\providecommand \@url [1]{\endgroup\@href {#1}{\urlprefix }}%
\providecommand \urlprefix  [0]{URL }%
\providecommand \Eprint [0]{\href }%
\providecommand \doibase [0]{http://dx.doi.org/}%
\providecommand \selectlanguage [0]{\@gobble}%
\providecommand \bibinfo  [0]{\@secondoftwo}%
\providecommand \bibfield  [0]{\@secondoftwo}%
\providecommand \translation [1]{[#1]}%
\providecommand \BibitemOpen [0]{}%
\providecommand \bibitemStop [0]{}%
\providecommand \bibitemNoStop [0]{.\EOS\space}%
\providecommand \EOS [0]{\spacefactor3000\relax}%
\providecommand \BibitemShut  [1]{\csname bibitem#1\endcsname}%
\let\auto@bib@innerbib\@empty
\bibitem [{\citenamefont {Tomas-Rodriguez}\ and\ \citenamefont
  {Banks}(2010)}]{tomas-rodriguez_linear_2010}%
  \BibitemOpen
  \bibfield  {author} {\bibinfo {author} {\bibfnamefont {M.}~\bibnamefont
  {Tomas-Rodriguez}}\ and\ \bibinfo {author} {\bibfnamefont {S.~P.}\
  \bibnamefont {Banks}},\ }\href@noop {} {{\selectlanguage {english}\emph
  {\bibinfo {title} {Linear, Time-varying Approximations to Nonlinear Dynamical
  Systems: With Applications in Control and Optimization}}}}\ (\bibinfo
  {publisher} {Springer Science \& Business Media},\ \bibinfo {year} {2010})\
  Chap.~\bibinfo {chapter} {2}\BibitemShut {NoStop}%
\bibitem [{\citenamefont {Curtain}\ and\ \citenamefont
  {Zwart}(1995)}]{curtain_introduction_1995}%
  \BibitemOpen
  \bibfield  {author} {\bibinfo {author} {\bibfnamefont {R.~F.}\ \bibnamefont
  {Curtain}}\ and\ \bibinfo {author} {\bibfnamefont {H.}~\bibnamefont
  {Zwart}},\ }\href@noop {} {{\selectlanguage {english}\emph {\bibinfo {title}
  {An Introduction to Infinite-Dimensional Linear Systems Theory}}}}\ (\bibinfo
   {publisher} {Springer Science \& Business Media},\ \bibinfo {year}
  {1995})\BibitemShut {NoStop}%
\bibitem [{\citenamefont {Söderström}(2002)}]{soderstrom_discrete-time_2002}%
  \BibitemOpen
  \bibfield  {author} {\bibinfo {author} {\bibfnamefont {T.}~\bibnamefont
  {Söderström}},\ }\href@noop {} {{\selectlanguage {english}\emph {\bibinfo
  {title} {Discrete-time Stochastic Systems: Estimation and Control}}}}\
  (\bibinfo  {publisher} {Springer Science \& Business Media},\ \bibinfo {year}
  {2002})\BibitemShut {NoStop}%
\bibitem [{\citenamefont {Steinbrecher}\ and\ \citenamefont
  {Weyssow}(2004)}]{steinbrecher_generalized_2004}%
  \BibitemOpen
  \bibfield  {author} {\bibinfo {author} {\bibfnamefont {G.}~\bibnamefont
  {Steinbrecher}}\ and\ \bibinfo {author} {\bibfnamefont {B.}~\bibnamefont
  {Weyssow}},\ }\href {\doibase 10.1103/PhysRevLett.92.125003} {\bibfield
  {journal} {\bibinfo  {journal} {Physical Review Letters}\ }\textbf {\bibinfo
  {volume} {92}},\ \bibinfo {pages} {125003} (\bibinfo {year}
  {2004})}\BibitemShut {NoStop}%
\bibitem [{\citenamefont {Eldar}\ and\ \citenamefont
  {Elowitz}(2010)}]{eldar_functional_2010}%
  \BibitemOpen
  \bibfield  {author} {\bibinfo {author} {\bibfnamefont {A.}~\bibnamefont
  {Eldar}}\ and\ \bibinfo {author} {\bibfnamefont {M.~B.}\ \bibnamefont
  {Elowitz}},\ }\href {\doibase 10.1038/nature09326} {\bibfield  {journal}
  {\bibinfo  {journal} {Nature}\ }\textbf {\bibinfo {volume} {467}},\ \bibinfo
  {pages} {167} (\bibinfo {year} {2010})}\BibitemShut {NoStop}%
\bibitem [{\citenamefont {Silver}(2010)}]{silver_neuronal_2010}%
  \BibitemOpen
  \bibfield  {author} {\bibinfo {author} {\bibfnamefont {R.~A.}\ \bibnamefont
  {Silver}},\ }\href {\doibase 10.1038/nrn2864} {\bibfield  {journal} {\bibinfo
   {journal} {Nature Reviews Neuroscience}\ }\textbf {\bibinfo {volume} {11}},\
  \bibinfo {pages} {474} (\bibinfo {year} {2010})}\BibitemShut {NoStop}%
\bibitem [{\citenamefont {Assaf}\ \emph {et~al.}(2013)\citenamefont {Assaf},
  \citenamefont {Roberts}, \citenamefont {Luthey-Schulten},\ and\ \citenamefont
  {Goldenfeld}}]{assaf_extrinsic_2013}%
  \BibitemOpen
  \bibfield  {author} {\bibinfo {author} {\bibfnamefont {M.}~\bibnamefont
  {Assaf}}, \bibinfo {author} {\bibfnamefont {E.}~\bibnamefont {Roberts}},
  \bibinfo {author} {\bibfnamefont {Z.}~\bibnamefont {Luthey-Schulten}}, \ and\
  \bibinfo {author} {\bibfnamefont {N.}~\bibnamefont {Goldenfeld}},\ }\href
  {\doibase 10.1103/PhysRevLett.111.058102} {\bibfield  {journal} {\bibinfo
  {journal} {Physical Review Letters}\ }\textbf {\bibinfo {volume} {111}},\
  \bibinfo {pages} {058102} (\bibinfo {year} {2013})}\BibitemShut {NoStop}%
\bibitem [{\citenamefont {Plerou}\ \emph {et~al.}(2002)\citenamefont {Plerou},
  \citenamefont {Gopikrishnan}, \citenamefont {Rosenow}, \citenamefont
  {Amaral}, \citenamefont {Guhr},\ and\ \citenamefont
  {Stanley}}]{plerou_random_2002}%
  \BibitemOpen
  \bibfield  {author} {\bibinfo {author} {\bibfnamefont {V.}~\bibnamefont
  {Plerou}}, \bibinfo {author} {\bibfnamefont {P.}~\bibnamefont
  {Gopikrishnan}}, \bibinfo {author} {\bibfnamefont {B.}~\bibnamefont
  {Rosenow}}, \bibinfo {author} {\bibfnamefont {L.~A.~N.}\ \bibnamefont
  {Amaral}}, \bibinfo {author} {\bibfnamefont {T.}~\bibnamefont {Guhr}}, \ and\
  \bibinfo {author} {\bibfnamefont {H.~E.}\ \bibnamefont {Stanley}},\ }\href
  {\doibase 10.1103/PhysRevE.65.066126} {\bibfield  {journal} {\bibinfo
  {journal} {Physical Review E}\ }\textbf {\bibinfo {volume} {65}},\ \bibinfo
  {pages} {066126} (\bibinfo {year} {2002})}\BibitemShut {NoStop}%
\bibitem [{\citenamefont {Hagin}(2004)}]{hagin_investment_2004}%
  \BibitemOpen
  \bibfield  {author} {\bibinfo {author} {\bibfnamefont {R.~L.}\ \bibnamefont
  {Hagin}},\ }\href@noop {} {{\selectlanguage {english}\emph {\bibinfo {title}
  {Investment Management: Portfolio Diversification, Risk, and Timing--Fact and
  Fiction}}}}\ (\bibinfo  {publisher} {John Wiley \& Sons},\ \bibinfo {year}
  {2004})\BibitemShut {NoStop}%
\bibitem [{\citenamefont {Bera}\ and\ \citenamefont
  {Park}(2008)}]{bera_optimal_2008}%
  \BibitemOpen
  \bibfield  {author} {\bibinfo {author} {\bibfnamefont {A.~K.}\ \bibnamefont
  {Bera}}\ and\ \bibinfo {author} {\bibfnamefont {S.~Y.}\ \bibnamefont
  {Park}},\ }\href {\doibase 10.1080/07474930801960394} {\bibfield  {journal}
  {\bibinfo  {journal} {Econometric Reviews}\ }\textbf {\bibinfo {volume}
  {27}},\ \bibinfo {pages} {484} (\bibinfo {year} {2008})}\BibitemShut
  {NoStop}%
\bibitem [{\citenamefont {Wang}\ \emph {et~al.}(2009)\citenamefont {Wang},
  \citenamefont {Yamasaki}, \citenamefont {Havlin},\ and\ \citenamefont
  {Stanley}}]{wang_multifactor_2009}%
  \BibitemOpen
  \bibfield  {author} {\bibinfo {author} {\bibfnamefont {F.}~\bibnamefont
  {Wang}}, \bibinfo {author} {\bibfnamefont {K.}~\bibnamefont {Yamasaki}},
  \bibinfo {author} {\bibfnamefont {S.}~\bibnamefont {Havlin}}, \ and\ \bibinfo
  {author} {\bibfnamefont {H.~E.}\ \bibnamefont {Stanley}},\ }\href {\doibase
  10.1103/PhysRevE.79.016103} {\bibfield  {journal} {\bibinfo  {journal}
  {Physical Review E}\ }\textbf {\bibinfo {volume} {79}},\ \bibinfo {pages}
  {016103} (\bibinfo {year} {2009})}\BibitemShut {NoStop}%
\bibitem [{\citenamefont {Kenny}(2009)}]{kenny_diversification_2009}%
  \BibitemOpen
  \bibfield  {author} {\bibinfo {author} {\bibfnamefont {G.}~\bibnamefont
  {Kenny}},\ }\href@noop {} {{\selectlanguage {english}\emph {\bibinfo {title}
  {Diversification Strategy: How to Grow a Business by Diversifying
  Successfully}}}}\ (\bibinfo  {publisher} {Kogan Page Publishers},\ \bibinfo
  {year} {2009})\BibitemShut {NoStop}%
\bibitem [{\citenamefont {Lande}\ and\ \citenamefont
  {Shannon}(1996)}]{lande_role_1996}%
  \BibitemOpen
  \bibfield  {author} {\bibinfo {author} {\bibfnamefont {R.}~\bibnamefont
  {Lande}}\ and\ \bibinfo {author} {\bibfnamefont {S.}~\bibnamefont
  {Shannon}},\ }\href {\doibase 10.2307/2410812} {\bibfield  {journal}
  {\bibinfo  {journal} {Evolution}\ }\textbf {\bibinfo {volume} {50}},\
  \bibinfo {pages} {434} (\bibinfo {year} {1996})}\BibitemShut {NoStop}%
\bibitem [{\citenamefont {Elena}\ and\ \citenamefont
  {Lenski}(2003)}]{elena_evolution_2003}%
  \BibitemOpen
  \bibfield  {author} {\bibinfo {author} {\bibfnamefont {S.~F.}\ \bibnamefont
  {Elena}}\ and\ \bibinfo {author} {\bibfnamefont {R.~E.}\ \bibnamefont
  {Lenski}},\ }\href {\doibase 10.1038/nrg1088} {\bibfield  {journal} {\bibinfo
   {journal} {Nature Reviews Genetics}\ }\textbf {\bibinfo {volume} {4}},\
  \bibinfo {pages} {457} (\bibinfo {year} {2003})}\BibitemShut {NoStop}%
\bibitem [{\citenamefont {de~Aguiar}\ and\ \citenamefont
  {Bar-Yam}(2011)}]{de_aguiar_moran_2011}%
  \BibitemOpen
  \bibfield  {author} {\bibinfo {author} {\bibfnamefont {M.~A.~M.}\
  \bibnamefont {de~Aguiar}}\ and\ \bibinfo {author} {\bibfnamefont
  {Y.}~\bibnamefont {Bar-Yam}},\ }\href {\doibase 10.1103/PhysRevE.84.031901}
  {\bibfield  {journal} {\bibinfo  {journal} {Physical Review E}\ }\textbf
  {\bibinfo {volume} {84}},\ \bibinfo {pages} {031901} (\bibinfo {year}
  {2011})}\BibitemShut {NoStop}%
\bibitem [{\citenamefont {Rulands}\ \emph {et~al.}(2014)\citenamefont
  {Rulands}, \citenamefont {Jahn},\ and\ \citenamefont
  {Frey}}]{rulands_specialization_2014}%
  \BibitemOpen
  \bibfield  {author} {\bibinfo {author} {\bibfnamefont {S.}~\bibnamefont
  {Rulands}}, \bibinfo {author} {\bibfnamefont {D.}~\bibnamefont {Jahn}}, \
  and\ \bibinfo {author} {\bibfnamefont {E.}~\bibnamefont {Frey}},\ }\href
  {\doibase 10.1103/PhysRevLett.113.108102} {\bibfield  {journal} {\bibinfo
  {journal} {Physical Review Letters}\ }\textbf {\bibinfo {volume} {113}},\
  \bibinfo {pages} {108102} (\bibinfo {year} {2014})}\BibitemShut {NoStop}%
\bibitem [{\citenamefont {Korolev}\ \emph {et~al.}(2010)\citenamefont
  {Korolev}, \citenamefont {Avlund}, \citenamefont {Hallatschek},\ and\
  \citenamefont {Nelson}}]{korolev_genetic_2010}%
  \BibitemOpen
  \bibfield  {author} {\bibinfo {author} {\bibfnamefont {K.~S.}\ \bibnamefont
  {Korolev}}, \bibinfo {author} {\bibfnamefont {M.}~\bibnamefont {Avlund}},
  \bibinfo {author} {\bibfnamefont {O.}~\bibnamefont {Hallatschek}}, \ and\
  \bibinfo {author} {\bibfnamefont {D.~R.}\ \bibnamefont {Nelson}},\ }\href
  {\doibase 10.1103/RevModPhys.82.1691} {\bibfield  {journal} {\bibinfo
  {journal} {Reviews of Modern Physics}\ }\textbf {\bibinfo {volume} {82}},\
  \bibinfo {pages} {1691} (\bibinfo {year} {2010})}\BibitemShut {NoStop}%
\bibitem [{\citenamefont {Meyer}\ and\ \citenamefont
  {Kassen}(2007)}]{meyer_effects_2007}%
  \BibitemOpen
  \bibfield  {author} {\bibinfo {author} {\bibfnamefont {J.~R.}\ \bibnamefont
  {Meyer}}\ and\ \bibinfo {author} {\bibfnamefont {R.}~\bibnamefont {Kassen}},\
  }\href {\doibase 10.1038/nature05599} {\bibfield  {journal} {\bibinfo
  {journal} {Nature}\ }\textbf {\bibinfo {volume} {446}},\ \bibinfo {pages}
  {432} (\bibinfo {year} {2007})}\BibitemShut {NoStop}%
\bibitem [{\citenamefont {Imbs}\ and\ \citenamefont
  {Wacziarg}(2003)}]{imbs_stages_2003}%
  \BibitemOpen
  \bibfield  {author} {\bibinfo {author} {\bibfnamefont {J.}~\bibnamefont
  {Imbs}}\ and\ \bibinfo {author} {\bibfnamefont {R.}~\bibnamefont
  {Wacziarg}},\ }\href {\doibase 10.1257/000282803321455160} {\bibfield
  {journal} {\bibinfo  {journal} {American Economic Review}\ }\textbf {\bibinfo
  {volume} {93}},\ \bibinfo {pages} {63} (\bibinfo {year} {2003})}\BibitemShut
  {NoStop}%
\bibitem [{\citenamefont {Peters}\ and\ \citenamefont
  {Klein}(2013)}]{peters_ergodicity_2013}%
  \BibitemOpen
  \bibfield  {author} {\bibinfo {author} {\bibfnamefont {O.}~\bibnamefont
  {Peters}}\ and\ \bibinfo {author} {\bibfnamefont {W.}~\bibnamefont {Klein}},\
  }\href {\doibase 10.1103/PhysRevLett.110.100603} {\bibfield  {journal}
  {\bibinfo  {journal} {Physical Review Letters}\ }\textbf {\bibinfo {volume}
  {110}},\ \bibinfo {pages} {100603} (\bibinfo {year} {2013})}\BibitemShut
  {NoStop}%
\bibitem [{\citenamefont {Porter}(2011)}]{porter_competitive_2011}%
  \BibitemOpen
  \bibfield  {author} {\bibinfo {author} {\bibfnamefont {M.~E.}\ \bibnamefont
  {Porter}},\ }\href@noop {} {{\selectlanguage {english}\emph {\bibinfo {title}
  {Competitive Advantage of Nations: Creating and Sustaining Superior
  Performance}}}}\ (\bibinfo  {publisher} {Simon and Schuster},\ \bibinfo
  {year} {2011})\BibitemShut {NoStop}%
\bibitem [{\citenamefont {Eigen}(1971)}]{eigen}%
  \BibitemOpen
  \bibfield  {author} {\bibinfo {author} {\bibfnamefont {M.}~\bibnamefont
  {Eigen}},\ }\href@noop {} {\bibfield  {journal} {\bibinfo  {journal}
  {Naturwissenschaften}\ }\textbf {\bibinfo {volume} {58}},\ \bibinfo {pages}
  {465} (\bibinfo {year} {1971})}\BibitemShut {NoStop}%
\bibitem [{\citenamefont {Tong}(2011)}]{tong_multivariate_2011}%
  \BibitemOpen
  \bibfield  {author} {\bibinfo {author} {\bibfnamefont {Y.~L.}\ \bibnamefont
  {Tong}},\ }\href@noop {} {{\selectlanguage {english}\emph {\bibinfo {title}
  {The Multivariate Normal Distribution}}}}\ (\bibinfo  {publisher} {Springer
  New York},\ \bibinfo {year} {2011})\BibitemShut {NoStop}%
\end{thebibliography}
\end{document}